\documentclass[twoside]{IEEEtran}

\usepackage{cite}
\usepackage{amsmath,amssymb,amsfonts}
\usepackage[shortlabels]{enumitem} 
\usepackage{graphicx}
\usepackage{color,soul} 
\newcommand\tred[1]{\textcolor[rgb]{0.98,0.00,0.00}{#1}} 

\newcommand\trev[1]{\textcolor[rgb]{0.00,0.00,0.00}{#1}} 
\newcommand\tdel[1]{\textcolor[rgb]{0.00,0.00,0.98}{\st{}}} 

\begin{document}
\bstctlcite{IEEEexample:BSTcontrol} 

\title{Asymmetric Si-Slot Coupler with Nonreciprocal Response Based on Graphene Saturable Absorption}

\author{Alexandros Pitilakis,
        Dimitrios Chatzidimitriou,
        Traianos Yioultsis, \IEEEmembership{Member, IEEE}, \\ and
        Emmanouil E. Kriezis, \IEEEmembership{Senior Member, IEEE}

\thanks{Manuscript received January 25, 2021; revised March 11, 2021; accepted March 31, 2021. This research is co-financed by Greece and the European Union (European Social Fund-ESF) through the Operational Programme ``Human Resources Development, Education and Lifelong Learning 2014-2020'' in the context of the project ``Design of nonlinear silicon devices incorporating graphene and using the Parity-Time symmetry concept'' (MIS 5047874). (\textit{Corresponding author: Alexandros Pitilakis.}) }
\thanks{All authors are with the Aristotle University of Thessaloniki, School of Electrical and Computer Engineering, 54124 Greece (email: alexpiti@auth.gr).}
\thanks{ \tred{\textcopyright~2021 IEEE. Personal use of this material is permitted, but republication/redistribution requires IEEE permission. Refer to IEEE Copyright and Publication Rights for more details.}}
\thanks{\tred{Digital Object Identifier (DOI): 10.1109/JQE.2021.3071247}}
\thanks{\tred{IEEE Xplore URL: https://ieeexplore.ieee.org/document/9395480}}
}

\markboth{IEEE Journal of Quantum Electronics | DOI: 10.1109/JQE.2021.3071247 }{Pitilakis \MakeLowercase{\textit{et al.}}}

\maketitle

\begin{abstract}
    We present the study of a proof-of-concept integrated device that can be used as a nonlinear broadband isolator. The device is based on the asymmetric loading of a highly-confining silicon-slot photonic coupler with graphene layers, whose ultrafast and low-threshold saturable absorption can be exploited for nonreciprocal transmission between the cross-ports of the coupler. The structure is essentially a non-Hermitian system, whose exceptional points are briefly discussed. The nonlinear device is modeled with a coupled Schr\"{o}dinger equation system whose validity is checked by full-vector finite element-based beam-propagation method simulations in CW. The numerically computed performance reveals a nonreciprocal intensity range (NRIR) in the vicinity of 100~mW peak power with a bandwidth spanning tens of nanometers, from CW down to ps-long pulses. Finally, the combination of saturable absorption and self-phase modulation (Kerr effect) in graphene is studied, indicating the existence of two NRIR with opposite directionality. 
\end{abstract}


\section{Introduction} \label{sec:1:intro}

The majority of passive and tunable photonic integrated circuits (PIC) and components are reciprocal, i.e., they exhibit exactly equal forward and backward transmission. Nonreciprocity is an often misunderstood \cite{Jalas2013,Caloz2018} electromagnetic (EM) property, denoting the absence of reciprocity, i.e., unequal transmission when input and output ports are interchanged. The archetype nonreciprocal component in guided-wave devices is the isolator, \tdel{the equivalent to the electric diode,} a two-port unidirectional device that allows low-loss forward transmission while blocking the backward one. The three-port extension of the isolator is a device with circular (azimuthal) symmetry which allows ``unirotational'' transmission between its ports, e.g., the input signal is only forwarded to the adjacent port in a fixed sense of rotation. Isolator and circulator functionalities are invaluable to source protection and full-duplex communication channels, respectively \cite{Sounas2017}. Specifically for optical communications, isolators are required to protect laser source cavities from destructive back-reflections, or to isolate parts of a circuit from harmful interference; similarly, circulators enable bi-directional communication over the same transmission channel, e.g., a single-mode fiber. Both functionalities are vital to optical transceivers, themselves essential to high-speed optical interconnects in datacenters, or emerging photonic applications such as LiDAR \cite{Yang2020} \trev{or sensors \cite{Xu2019}}. 

Fundamental EM theory allows three avenues to ``breaking'' reciprocity, i.e, time-reversal symmetry: (i) magnetic properties \cite{Dotsch2005}, (ii) space-time modulation \cite{Sounas2013}, or (iii) nonlinearity combined with spatial asymmetry \cite{Sounas2018}\tdel{\nocite{Antonellis2019}}. The present work focuses in the latter, which does not require active elements or multiple waves (unlike space-time modulation) and does not implicate magneto-optic materials which are bulky and incompatible with contemporary PIC technologies, e.g. SOI (silicon-on-insulator) or SiN (silicon nitride), with few exceptions \cite{Huang2016}. \trev{Nonreciprocity through nonlinearity, see Section XXI in \cite{Caloz2018}, additionally requires for spatial asymmetry in the structure; moreover, nonlinear isolators are subject to inherent bounds such as limited range of powers, half-duplex operation in CW (i.e., simultaneous excitation from both directions is prohibited), or asymptotic performance thresholds. Partially overcoming these limitations, and building upon expertise in nonlinear graphene-comprising \cite{Chatzidimitriou:2015,Chatzidimitriou:18,Chatzidimitriou:20} and hybrid silicon photonic design \cite{Pitilakis:2013,Tsilipakos2011},} we demonstrate a proof-of-concept device based on graphene \textit{saturable absorption} (SA) in a non-resonant structure operating in the NIR (1550~nm) region. Our device is an asymmetrically-loaded SOI directional coupler, where the loading consists of graphene sheets \cite{Ferrari2015,Chamanara2013}, motivated by the broadband response and the rather low SA intensity-threshold \cite{Bao2009,Marini2016}. The technological maturity of the SOI platform is indispensable in engineering tightly confining graphene-loaded waveguides, so as to maximize the loss-contrast between the low- and high-power regime, simultaneously decreasing the power-threshold of SA-onset and eliminating unwanted nonlinear effects, e.g., from silicon. 

This device\tdel{, which can be considered as the photonic analogue of the Zener diode,} has three operation regimes: bidirectional isolation at low powers, half-duplex isolation for powers inside the \textit{nonreciprocal intensity range} (NRIR), and bidirectional transmission above a higher ``breakdown'' power. Our approach deviates in two aspects from the more frequently used phase-related nonlinearities (e.g., Kerr effect) implemented in resonant cavities \cite{DelBino2018,Rodriguez2019,Yang2020}, offering half-duplex isolator performance in a very large bandwidth, and thus has potential applications in high-fluence fs-pulsed on-chip sources. Moreover, we offer a novel design concept, based on a non-Hermitian system, i.e., a pair of coupled subsystems with asymmetry in their loss, with its signature exceptional points delimiting sharp changes in their response \cite{Kominis2017,Chatzidimitriou:18,Miri:19}; note that a special class of non-Hermitian systems are those exhibiting parity-time ($\mathcal{PT}$) symmetry, where exactly balanced gain and loss are present. Finally, we note that graphene SA has potential applications in all-optical interconnects or pulsed-source components, e.g., as an SA mirror \cite{Mock2017}.

The remainder of this paper is organized as follows: Section~\ref{sec:2:concept} presents the device concept, physical description of the graphene SA used, and coupled-equation modeling of the non-Hermitian system. Section~\ref{sec:3:physical} contains the implementation in a graphene-clad Si-photonic waveguide coupler and its simulated CW performance. Section~\ref{sec:4:Further} addresses the pulsed regime performance and the combined effect of SA and Kerr effect. Section~\ref{sec:5:conclusions} provides the conclusions of our work.

\section{Device Concept and Framework} \label{sec:2:concept}
\subsection{Nonreciprocal Asymmetrically-loaded Coupler} \label{sec:ConceptDescription}

\begin{figure}[]
    \centering
    \includegraphics[]{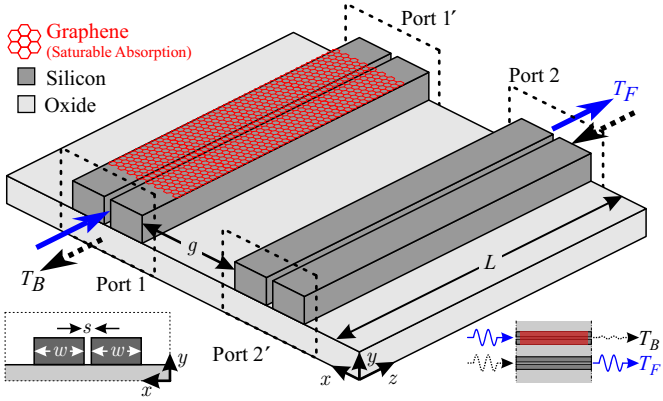}
    \caption{Schematic of the asymmetrically loaded Si-slot waveguide coupler \trev{with annotated dimensions; $xy$-axes are in-scale with $g\approx 1~\mu$m and $z$-length $L$ is a few hundred microns}. When used as a two-port nonreciprocal structure, the forward and backward transmission is defined between the ``cross'' ports of the coupler, i.e., $T_F=T_{2\leftarrow1}$ and $T_B=T_{1\leftarrow2}$. Bottom right-hand inset: Due to the symmetry in the structure, we can interchange primed and unprimed ports, and in all cases the unused ``bar'' output ports are assumed matched.}
    \label{fig:Schem1}    
\end{figure}

A schematic of the directional coupler is illustrated in Fig.~\ref{fig:Schem1}, where a graphene ribbon asymmetrically loads only one of the silicon-slot waveguides\trev{; the device $z$-length $L$ is a few hundred microns and the $x$-gap between the two slot waveguides is $g\approx 1~\mu$m. The nonreciprocal response is due to the SA in graphene and manifests as unequal forward and backward ``cross-port'' transmission, $T_F=T_{2\leftarrow1} \neq T_B=T_{1\leftarrow2}$. In the CW regime, only half-duplex isolation can be achieved, i.e., we can excite only one port at a time (1:forward, 2:backward);} full-duplex isolation is possible in the pulsed regime, provided that the pulse duration and repetition-rate are low. Note that the underlying photonic coupler in the absence of graphene loading is \textit{synchronized}, i.e., its two Si-slot waveguides are identical in all their geometric and EM parameters. Also, the structure is $z$-invariant and all ports are non-reflecting.

\begin{figure}[]
    \centering
    \includegraphics[]{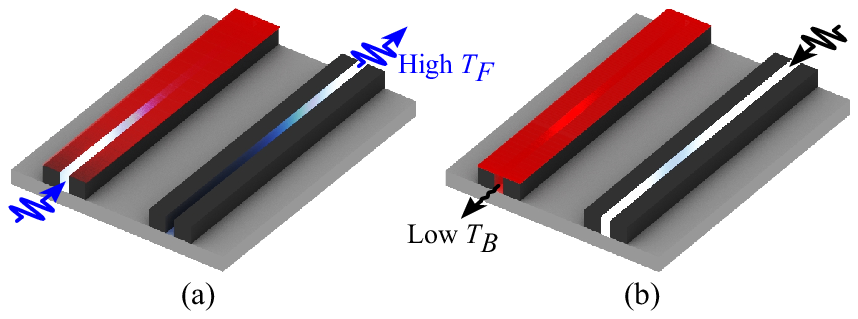}
    \caption{Concept illustration of the (a) high forward and (b) low backward transmission that can be attained for input powers inside the NRIR. The length of the asymmetrically-loaded nonlinear device is equal to the coupling-length of the underlying Si-slot waveguide coupler (in the absence of graphene). \trev{Indicative geometric dimensions can be found in Fig.~\ref{fig:Schem1}, \ref{fig:SingleWG_GeomOptim}, \ref{fig:SingleWG_LossSatComp}, and \ref{fig:Coupler_GeomOptim}}}
    \label{fig:Concept}    
\end{figure}

Assuming that the directional coupler $z$-length is approximately equal to the coupling length ($L_c$) of the device in the absence of the graphene-SA loading, the operation concept can be described as follows. In the low-power (linear) regime, the large asymmetry in the losses between the two waveguides means that coupling is inhibited, and cross-transmission is very low; in this regime the two-port device is reciprocal with very low transmission in both directions, \trev{$T_F\approx T_B\rightarrow0$}. Now, nonreciprocity is attained in the nonlinear regime, for input power inside the NRIR, which lies above the loaded-waveguide SA threshold. When exciting the graphene-loaded waveguide, the high power quenches its losses thanks to SA so that both waveguides are practically lossless and the coupler is almost synchronized; this allows the signal to cross to the lossless waveguide and this is the ``forward'' or through direction, with high transmission \trev{$T_F\rightarrow1$}, Fig.~\ref{fig:Concept}(a). On the contrary, when exciting the lossless waveguide with a moderately high power, the losses in the opposite (graphene-loaded) waveguide remain high so that cross-coupling is inhibited due to the asymmetry; this is the ``backward'' or isolated direction, with low transmission \trev{$T_B\rightarrow0$}, Fig.~\ref{fig:Concept}(b). Finally, when the backward excitation power exceeds a threshold value, cross-saturation synchronizes the coupler allowing high backward transmission; this is the ``breakdown'' regime of the device with quasi-reciprocal high-transmission in both directions, \trev{$T_F\approx T_B\rightarrow1$}. The asymmetry between the transmission in the two directions for powers inside the NRIR can be engineered in a half-duplex isolator, based on the nonlinearity and on the asymmetric graphene-loading.

\subsection{Saturable Absorption in Graphene} \label{sec:Graphene_SA}

As described in Section~\ref{sec:ConceptDescription}, the device operation relies on saturable absorption, i.e., the nonlinear quenching of losses with increasing power. In a perturbative third-order nonlinear regime, SA can be treated with a term similar to the one commonly used for two-photon absorption (TPA) but of opposite sign. TPA is a nonlinear process that increases the losses for high intensity signals, thus the sign reversal, and manifests in semiconductors by absorbing photons above half the bandgap energy and generating free carriers \cite{Lin2007}. In most materials, SA is typically observed at higher power thresholds, closer to optical damage, in which case it ceases to fall into the perturbative regime. The atoms absorbing the radiation energy are, in their majority, excited to higher energy states and can no longer efficiently relax their energy to the lattice so that they can be re-excited and absorb more energy. This process leads to SA and culminates in the breakdown of the material (irreversible damage) as power is further increased. The critical parameters for any SA material are the saturation intensity (in W/m$^2$, defined as the CW intensity for which absorption reduces to half of the low-power regime) or fluence (in J/m$^2$, for pulsed excitation), and the relaxation time, i.e., the time required for the material to desaturate, shedding its energy to the lattice as its atoms decay to lower energy states.

Graphene, a 2D semi-metal or zero bandgap semiconductor \cite{Ferrari2015}, can be cast as a high-contrast SA material in the NIR, owing to its mono-atomic thickness and the gap-less Dirac-cone dispersion. Various theoretical models have been proposed for its nonlinear behaviour, in perturbative \cite{Cheng2015,Mikhailov2016Quantum} and non-perturbative \cite{Marini2016,Marini2017,Mikhailov2019HEM} regimes, using semi-classical and/or thermodynamic tools. These models lead to standard third-order nonlinear response (Kerr effect, self/cross-phase modulation, four-wave mixing, parametric conversion) or to a more complicated response, when coupled to the photo-excited carrier plasma \cite{Alexander2018,Demongodin2019,CastellLurbe2020}. All these models predict a SA regime for graphene, when it is biased below the half-photon energy where interband electronic transitions are not restricted by Pauli blocking and, consequently, absorption is high (metallic regime). When biased above that threshold energy, graphene is practically transparent in the NIR due to the absence of interband mechanism (dielectric regime) and, moreover, it exhibits TPA \cite{Cheng2015,Mikhailov2016Quantum}. One can understand the SA behavior in simplistic thermodynamic terms as follows: In the loss regime, graphene carriers absorb the EM energy and their excitation leads to a nearly instantaneous (tens of fs timescale) heating; the effect on the surface conductivity of this large temperature increase is a blurring between the inter- and intraband mechanisms \cite{Cheng2015} and eventually a transition between its two regimes, the high loss (metallic) and low loss (dielectric). Desaturation happens at slower timescales, in the ps-order, due to interband recombination and various scattering processes \cite{Mikhailov2019HEM} \tdel{\nocite{Soavi2019}}. So, if the graphene Fermi energy was set within the bounds of the high-loss regime then high-intensity illumination will decrease the losses; the higher the intensity, the higher the saturation of losses and the higher the loss contrast, i.e., the difference in losses between the linear (low-power) and the nonlinear (high-power) regime. We note that this field is currently under intense investigation, with large deviations in the reported nonlinear parameters and the thresholds between perturbative/non-perturbative regimes. These aspects transcend the scope of this work, which is to investigate the performance of SA-enabled nonreciprocity in a realistic proof-of-concept device. In this spirit, we assume an instantaneous SA response with a phenomenological model for graphene conductivity and study its spatially averaged effect on the optical propagation in picosecond temporal regimes.

In such a model, the graphene conductivity can be separated in two parts, the non-saturable and the saturable, which are directly attributed to the intraband [$\sigma^{(1)}_i$] and interband [$\sigma^{(1)}_e$] mechanisms, respectively. The sum of these terms forms the total linear conductivity of graphene, $\sigma^{(1)}=\sigma^{(1)}_i+\sigma^{(1)}_e$, and depends on its effective chemical potential (assumed fixed and below the half-photon energy, $|\mu_c|<\hbar\omega/2$) and its temperature (assumed fixed at equilibrium, $T=300$~K); exact expressions can be found in \cite{Chatzidimitriou:2015}. The non-saturable part is independent of the incident radiation whereas the saturable part is assumed to scale with the phenomenological law $1/(1+\rho)$, with $\rho$ being proportional to the optical intensity; the linear and SA regimes are denoted by $\rho\ll1$ and $\rho>1$, respectively. In this work $\rho=|\mathbf{E}_\parallel|^2/E_\mathrm{sat}^2$, where $\mathbf{E}_\parallel$ is the E-field component parallel to the graphene sheet(s), $E_\mathrm{sat}^2=2 Z_0 I_\mathrm{sat}$, $Z_0=377$~Ohm, and $I_\mathrm{sat}$ is the saturation intensity. For the latter, we use the value $I_\mathrm{sat}=1$~MW/cm$^2$ \cite{Bao2009,Zhang2015}. In terms of our full-wave EM simulations the ``effective'' surface conductivity across the structure is
    \begin{equation}
        \sigma^{(1)}(x,y,z) = \sigma^{(1)}_i + \sigma^{(1)}_e \frac{1}{1+|\mathbf{E_\parallel}(x,y,z)|^2/E_\mathrm{sat}^2},
    \label{eq:sigma1SA}
    \end{equation}
where $\sigma^{(1)}_{i,e}$ are constants, assuming uniform graphene sheets (fixed $\mu_c$, $T$ and $\omega$). Consequently, the macroscopic spatial inhomogeneity in this effective $\sigma^{(1)}$ depends only on the local E-field intensity, and is thus nonlinear. To further simplify our analysis and concept implementation, focusing on the upper performance threshold, we assume that since graphene is biased below the half-photon energy, the interband conductivity dominates [$\sigma^{(1)}_i\approx0$] and it moreover acquires a real constant value, i.e., $\sigma^{(1)}_e\approx \sigma_0=e^2/4\hbar\approx61~\mu$S, where $\sigma_0$ is the ``universal'' optical conductivity of graphene responsible for the 2.3\% absorption through an air-suspended monolayer.

Note that, as high-confinement waveguides support hybrid modes, we account for the tensor properties of the 2D material. In this sense, the value of \eqref{eq:sigma1SA} corresponds to both nonzero elements of the main diagonal of the second-rank tensor describing graphene \tdel{. For instance, if graphene lies in the $xz$-plane, then $\sigma^{(1)}_{xx}=\sigma^{(1)}_{zz}=\sigma^{(1)}$, $\sigma^{(1)}_{yy}=0$ and $\sigma^{(1)}_{ab}=0$ for $a \neq b$, which describes} as an isotropic 2D material \cite{Chatzidimitriou:2015}. 

\subsection{Coupled Mode Framework} \label{sec:Coupled_NLSE}

For the mathematical modeling of light propagation in this nonreciprocal device, we employ a coupled-mode theory approach, specifically, a pair of coupled nonlinear Schr\"{o}dinger equations (NLSE). This framework properly accounts for the waveguide geometry and the linear/nonlinear macroscopic response of constituting materials on the spatial distribution of the guided modes, through effective parameters rigorously calculated for the specific physical implementation.

One of the prerequisites for the NLSE derivation and validity is that the spatial eigenmode profiles are unaltered during propagation, which is the case for perturbative nonlinearity (Kerr effect) in multimode single-core waveguides, such as birefringent fibers. As long as all guided modes are only slightly perturbed during propagation, this concept can be extended to multi-core waveguides such as directional couplers \cite{Pitilakis:2013}, where we have the ``supermodes'', i.e., eigenmodes with symmetric and anti-symmetric profiles, in the synchronized case. Introducing non-perturbative nonlinearity to the coupler will force the coupler eigenmodes to substantially change along the propagation and their evolution will moreover depend on the symmetry (or lack) of the initial excitation. Additionally, if the structure is asymmetric with respect to the material absorption, then we have a non-Hermitian system with exceptional points (EP), \cite{Chatzidimitriou:18}, which non-trivially affect the eigenmode profiles. Specifically, in the asymmetric SA-loaded directional coupler structure, there is one EP that can be identified as the SA level where the two eigenmodes of the structure \textit{coalesce}, i.e., when the eigenvalues \textit{and} mode profiles converge; this mode coalescence must not be confused with mirror symmetry or degeneracy, as we are considering asymmetric single-polarization waveguides. The combination of these modifications (non-perturbative nonlinear loss-asymmetry), render the coupled-supermode NLSE framework unusable, because crossing the EP imparts a substantial change in the mode profiles during propagation.  

To overcome this obstacle, we assume that the modification of the coupler eigenmodes is almost solely attributed to the non-Hermitian nature of the system and not to the modification of the underlying material EM properties. In other words, nonlinearity does not substantially modify the mode profiles of the \textit{isolated} waveguides. This assumption holds true for NIR waveguides comprising graphene sheets, which are not contributing to mode confinement or guidance. Thus, in this work we derive two separate NLSEs, one for each isolated waveguide of the coupler. We then couple the two equations with a coefficient derived in the linear regime and when the asymmetry (the graphene loading) is absent. This approach implies that only ``self-acting'' nonlinear effects (such as self-SA and Kerr) are considered and phenomena like direct cross-phase/amplitude modulation are negligible. This approximation is valid in the weak-coupling regime that we are considering, as will be demonstrated in Section~\ref{sec:3:3:CW} by means of numerical simulations. Do note, however, that \textit{indirect} cross-effects such as cross-absorption modulation are still allowed, as power is exchanged between the waveguides.

The derivation of the NLSE is a subject extensively covered in literature, e.g., in \cite{Lin2007,Pitilakis:2013}. Here, we directly present the general form of the loosely coupled NLSE system, under the $e^{+j\omega t}$ harmonic oscillation phase convention,
    \begin{equation}
    \label{eq:NLSE_general}
        \dfrac{\partial}{\partial z}
            \begin{bmatrix}
                A_1\\
                A_2
            \end{bmatrix}
        =
        \begin{bmatrix}
             +\delta^{(1)} & -j\kappa \\ 
             -j\kappa      & +\delta^{(2)}
        \end{bmatrix}
        \begin{bmatrix}
        A_1\\
        A_2
        \end{bmatrix}
    ,
    \end{equation}
where $A_k=A_k(z,\tau)$ are the complex amplitudes of the guided mode envelopes in the $k=\{1,2\}$ waveguide (e.g., the loaded and unloaded waveguides in Fig.~\ref{fig:Schem1}, respectively) measured in W$^{1/2}$, $\kappa=\pi/(2L_c)$ the coupling coefficient, and $\delta^{(k)}$ the $k$-th mode ``self-acting'' term:
    \begin{equation}
        \label{eq:NLSE_selfact_term}
        \delta^{(k)} = -\frac{\alpha^{(k)}}{2} + j\Delta\beta_\mathrm{NL}^{(k)} + j\gamma^{(k)}|A_k|^2 + D^{(k)} .
    \end{equation}
In this compact term, $\alpha^{(k)}$ is the power loss/gain coefficients (if positive/negative, respectively), $\gamma^{(k)}$ the complex third-order nonlinear parameter (including Kerr effect and perturbative SA/TPA), and $\Delta\beta_\mathrm{NL}^{(k)}$ includes nonlinear phase-dispersion contributions excluding third-order effects which are included in $\gamma^{(k)}$. $D^{(k)}$ is the linear dispersion operator,
    \begin{equation}
        \label{eq:NLSE_Dispersion}
        D^{(k)} = \left(\frac{1}{\overline{v}_\mathrm{g}}-\frac{1}{v^{(k)}_\mathrm{g}}\right) \frac{\partial}{\partial \tau}
         +\sum_{m=2}^{\infty}(-j)^{m+1}\frac{\beta_m^{(k)}}{m!}\frac{\partial^m }{\partial \tau^m},
    \end{equation}
where $\overline{v}_\mathrm{g}=(v^{(1)}_\mathrm{g}+v^{(2)}_\mathrm{g})/2$ is the mean group velocity, $v^{(k)}_\mathrm{g}$ the group velocity, $\beta_m^{(k)}$ are the $m$-th dispersion parameters ($m=2,3$ is group-velocity dispersion, GVD, and third-order dispersion, TOD, respectively), and $\tau$ is a retarded time frame, moving with $\overline{v}_\mathrm{g}$. All parameters in \eqref{eq:NLSE_general}, \eqref{eq:NLSE_selfact_term} and \eqref{eq:NLSE_Dispersion} are evaluated at a central frequency $\omega_0$, and all are real-valued unless explicitly stated. Finally, note that $\Delta\beta_\mathrm{NL}^{(k)}$ and $\alpha^{(k)}$, are allowed only ``self-acting'' nonlinearity, i.e., they exclusively depend on $A_k(z,\tau)$. This models effects that do not fall into standard categories (these being the higher-order dispersion terms and the third-order effects, modeled by $D^{(k)}$ and $\gamma^{(k)}$, respectively), such as non-perturbative SA or saturable photo-generated carrier refraction \cite{Vermeulen2018,CastellLurbe2020}. In the latter case, an additional rate equation is required, coupled to the NLSE system through $\Delta\beta_\mathrm{NL}^{(k)}$ and/or $\alpha^{(k)}$, which governs the temporal dynamics of the free-carrier plasma generated by the optical envelope \cite{Alexander2018}.

The coupling coefficient ($\kappa$) is the only parameter evaluated from the coupler as a whole and not from the individual waveguides. Specifically, if $\beta_\mathrm{S}$ and $\beta_\mathrm{A}$ are the phase constants of the symmetric and anti-symmetric supermodes, respectively, then the coupling length is given by $L_c=\pi/(\beta_\mathrm{S}-\beta_\mathrm{A})=\pi/2\kappa$. The frequency dispersion of $\kappa$ can be added to the coupled system in the frequency domain, either with a Taylor series expansion around $\omega_0$ or directly, from its spectrum $\kappa(\omega)$.

\begin{figure}[b]
    \centering
    \includegraphics[]{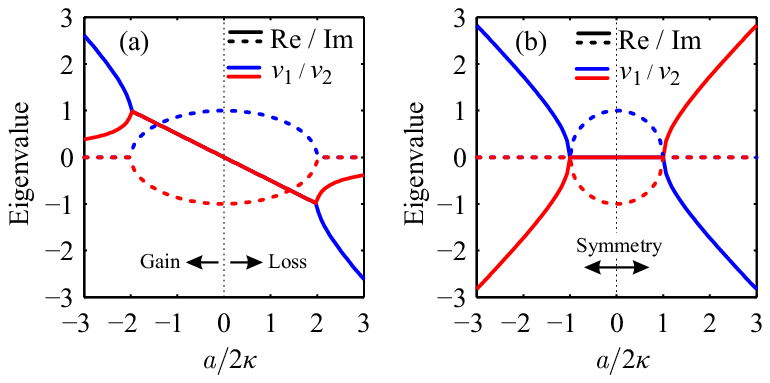}
    \caption{Evolution of eigenvalues in a non-Hermitian system. (a) Asymmetric coupler with one transparent waveguide and one waveguide with loss or, hypothetically, gain. (b) $\mathcal{PT}$-symmetric coupler, with exactly equal loss and gain in each of its waveguides, not further studied in this work.}
    \label{fig:EPs}    
\end{figure}

To gain insight into the non-Hermitian system evolution, \eqref{eq:NLSE_selfact_term} can be cast in a much simpler form, including only the parameters relevant to our asymmetrically SA-loaded coupler. Specifically, assuming the CW regime (where the time-derivatives vanish), absence of third-order nonlinearity, and the implications of our instantaneous SA model presented in Section~\ref{sec:Graphene_SA}, the coupled equation system for the asymmetric coupler can be dramatically simplified to
\begin{equation}
\label{eq:CMT_simple}
    \dfrac{\partial}{\partial z}
        \begin{bmatrix}
            A_1\\
            A_2
        \end{bmatrix}
    =
    \begin{bmatrix}
         -\alpha/2 & -j\kappa \\
         -j\kappa  & 0
    \end{bmatrix}
    \begin{bmatrix}
    A_1\\
    A_2
    \end{bmatrix}
,
\end{equation}
which entails only the coupling coefficient, $\kappa$, and the power-attenuation factor in the first waveguide, $\alpha=\alpha(|A_1|^2)$, where we have dropped the superscript. The latter describes the \textit{saturation curve} of the graphene-loaded waveguide, having a high value ($\alpha_0$) at low powers and a monotonic decrease as the power decreases; the saturation power ($P_\mathrm{sat}$) is defined as the value of $|A_1|^2$ for which $\alpha=\alpha_0/2$. The two eigenvalues $\nu_{1,2}$ of the system can be easily computed from the matrix in \eqref{eq:CMT_simple} as a function of the normalized parameter \trev{$\alpha/2\kappa$}. This unveils the EP at \trev{$\alpha/2\kappa=2$} where modes coalescence, Fig.~\ref{fig:EPs}(a), as well as the hypothetical case of a gain factor, which has a symmetric EP at \trev{$\alpha/2\kappa=-2$}. It is worth depicting the eigenvalues of the $\mathcal{PT}$-symmetric case, i.e., the special case of a non-Hermitian system with exactly matched gain and loss in each of the coupler waveguides, Fig.~\ref{fig:EPs}(b), whose EPs lie at \trev{$\alpha/2\kappa=\pm1$}. More detailed discussions on the nuances and potential applications these features can be found in \cite{Chatzidimitriou:18,Kominis2017}.

\section{Physical Implementation and CW Performance} \label{sec:3:physical}
\subsection{Waveguide and Coupler Design} \label{sec:3:Waveguide}

\begin{figure}[] 
    \centering
    \includegraphics[]{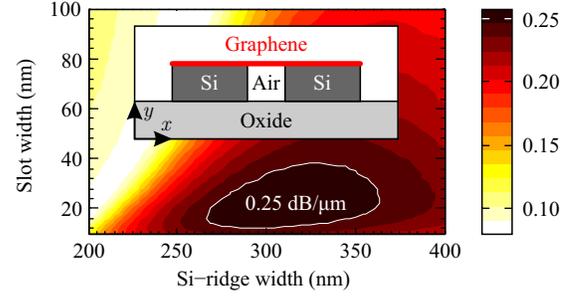}
    \caption{Propagation losses (dB/$\mu$m) as a function of the waveguide cross-section, depicted in the inset. The silicon ridge height is 140~nm and the graphene ribbon (thick red line in the inset) has a uniform $\sigma^{(1)}=61~\mu$S. }
    \label{fig:SingleWG_GeomOptim}
\end{figure}

In order to enhance the light-matter interaction in our structure, and so boost the nonlinear effects originating from the 2D material (graphene), we select the slot waveguide geometry, where light is confined in a low index material (air) between two high-index ridges (silicon). The waveguide cross-section is depicted in the inset of Fig.~\ref{fig:SingleWG_GeomOptim}, characterized by high confinement as the slot width is reduced. This applies to the quasi-TE mode, having a horizontally polarized transverse E-field component, so that it is parallel to a 2D material patterned in a ribbon and extending out to the outer vertical walls of the Si-ridges. In order to optimize the waveguide dimensions, i.e., the Si-ridge height and width, and the slot width, we employ a finite element method (FEM) based eigenmode solver \cite{Selleri2001} and extract the modal attenuation factor. As the absorption in the waveguide is exclusively due to graphene conductivity ($\mathrm{Re}\{\sigma^{(1)}\}$), we seek the geometric dimensions that maximize the modal power-loss constant, $\alpha$. If graphene sheets are included in the eigenmode problem, then $\alpha=-2 k_0 \mathrm{Im}\{n_\mathrm{eff}\}$, where $n_\mathrm{eff}$ is the complex effective index of the eigenmode. Note that, as graphene does not contribute to the waveguiding in the NIR\footnotemark, i.e., $|\mathrm{Im}\{\sigma^{(1)}\}|\approx \omega\varepsilon_0 d_\mathrm{gr}$ is in the few-$\mu$S region, and as the 2D material tensor is isotropic, we can accurately estimate the waveguide losses perturbatively:
\begin{equation}
    \alpha = \frac{1}{2\mathcal{P}} \int_G \sigma^{(1)}(x,y)|\mathbf{e}_\parallel(x,y)|^2 \mathrm{d}\ell. 
\label{eq:LossesWG}
\end{equation}
In this expression, vector $\mathbf{e}(x,y)$ is the eigenmode profile extracted by the solver in the absence of graphene-loading, the line-integral is performed in the waveguide cross-section assumed to be occupied by graphene sheets, $\sigma^{(1)}(x,y)\neq0$, and it uses only the E-field components that are parallel to graphene ($\mathbf{e}_\parallel$). The scalar $\mathcal{P}=0.5\iint\mathrm{Re}\{\mathbf{e}\times\mathbf{h}^*\}\cdot \hat{\mathbf{z}}\mathrm{d}x\mathrm{d}y$ is an eigenmode-dependent normalization constant, in Watt.

\footnotetext{As $\varepsilon_{r,\mathrm{eff}}=1-j \sigma^{(1)}/(\omega\varepsilon_0 d_\mathrm{gr})$, $d_\mathrm{gr}=0.35$~nm the effective thickness of a graphene monolayer, in the FIR/THz region the large negative $\mathrm{Im}\{\sigma^{(1)}\}$ leads to $\mathrm{Re}\{\varepsilon_{r,\mathrm{eff}}\}\ll-1$ which, in turn, gives rise to plasmonic waveguiding, i.e., strong confinement of the E-field perpendicular to graphene.}

Assuming a finite-width graphene ribbon (patterned to cover the air-slot and the two Si-ridges) of uniform $\sigma^{(1)}=61~\mu$S, we numerically optimize the geometric parameters seeking for a maximization of the propagation losses for the $x$-polarized mode at $\lambda_0=1550$~nm. The oxide and silicon refractive indices were $n_\mathrm{Ox}=1.45$ $n_\mathrm{Si}=3.47$, respectively. We noted that the optimal silicon-ridge height is below 200~nm, so that the E-field in the upper part of the air-slot can sufficiently overlap with graphene, and as close as possible to the cut-off thickness where the mode leaks into the oxide substrate. Fixing the height at the technologically acceptable value of 140~nm, we calculate the optimal combination of Si-ridge width and slot width, 300~nm and 20~nm, respectively, depicted in Fig.~\ref{fig:SingleWG_GeomOptim}. The maximum propagation loss is almost 0.27~dB/$\mu$m with a reasonable tolerance on the geometric parameters, ensuring a low fabrication sensitivity in this design. Note that for Si-ridge widths above 300~nm the waveguide also supports a low-loss anti-symmetric $x$-polarized mode, localized inside the silicon cores, which is unwanted. If the monolayer is replaced by few-layer graphene, then the absorption is expected to increase proportionally to the number of layers, as long as the layers are assumed uncoupled and sub-$\mathrm{nm}$ thick in total; for instance, an uncoupled bilayer ribbon will have $\sigma^{(1)}=122~\mu$S which will lead to 0.54~dB/$\mu$m losses. Finally, we note that selecting an \textit{infinite}-width graphene monolayer sheet instead of a ribbon, would slightly increase the losses, up to 0.31~dB/$\mu$m, owing to the E-field concentrated in the upper/outer corners of the Si-ridges. We nevertheless opt for the ribbon design as we anticipate that it would limit the diffusion of photo-excited carriers in graphene, which reduces the local carrier density and consequently increases the saturation intensity of the waveguide \cite{Chatzidimitriou:20}, an unwanted effect in our device.

\begin{figure}[] 
    \centering
    \includegraphics[]{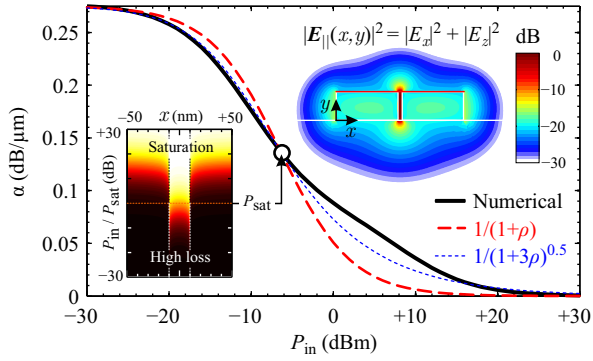}
    \caption{Nonlinear loss-saturation curve for the optimized graphene monolayer-loaded waveguide, compared to phenomenological models where $\rho=P_\mathrm{in}/P_\mathrm{sat}$. For the selected waveguide design, $P_\mathrm{sat}\approx-6$~dBm is where the waveguide losses are halved with respect to the low-power (linear) regime. The right inset depicts the $|\mathbf{E}_\parallel|^2$ profile in the cross-section. The left inset depicts the saturation of graphene's local conductivity over the slot region (horizontal axis) as the input power increases (vertical axis), with dark and light colors denoting absorptive and transparent regions, respectively. }
    \label{fig:SingleWG_LossSatComp}
\end{figure}

Having selected the Si-slot waveguide geometric parameters, we can estimate the SA curve in the waveguide, i.e., how the losses ($\alpha$) depend on the CW power launched into the mode ($P_\mathrm{in}$). We use the waveguide mode profile in the absence of graphene with the approximation of \eqref{eq:LossesWG}, where now $\sigma^{(1)}$ is power-dependent, i.e., as in \eqref{eq:sigma1SA} with $|\mathbf{E_\parallel}(x,y)|^2 \longrightarrow (P_\mathrm{in}/\mathcal{P}) |\mathbf{e_\parallel}(x,y)|^2$; for the uniform graphene ribbon we assume $\sigma^{(1)}_i=0$ and $\sigma^{(1)}_e=61$~$\mu$S. The resulting loss-saturation curve is depicted in Fig.~\ref{fig:SingleWG_LossSatComp}, with a thick black line, from which we extract a sub-mW saturation power of $P_\mathrm{sat}\approx-6$~dBm or 0.22~mW. We also compare the numerically calculated curve with commonly used phenomenological models $\alpha = \alpha_0/(1+\rho)$ and $\alpha = \alpha_0/\sqrt{1+3\rho}$, where $\rho=P_\mathrm{in}/P_\mathrm{sat}$ and $\alpha_0$ are the low-power losses. While all models qualitatively agree below or close to $P_\mathrm{sat}$, the deviations become non-negligible at higher powers which is expected to influence the component performance. The two insets in Fig.~\ref{fig:SingleWG_LossSatComp} depict the very high confinement of the $xz$-polarized E-field components inside the slot, leading to a deep saturation of graphene conductivity in its vicinity, even at modest power levels.

\begin{figure}[] 
    \centering
    \includegraphics[]{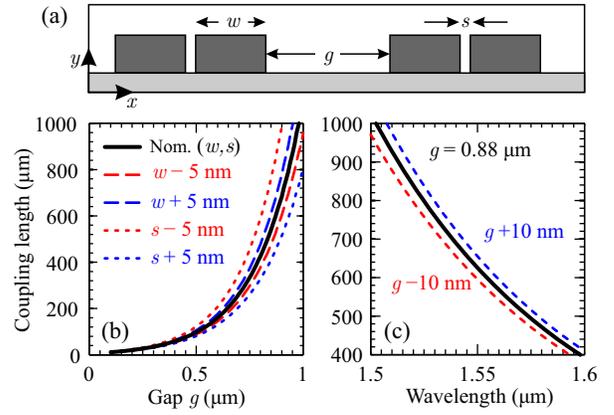}
    \caption{(a) Cross section of the symmetric (unloaded) Si-slot waveguide coupler and primary dimensions. (b) Coupling length vs. gap for $w=300$~nm and $s=20$~nm, also presenting the deviation for few-nm sized offsets from these nominal parameters. (c) Wavelength dispersion of the coupling length around $1550$~nm for three gap values spaced by 10~nm.}
    \label{fig:Coupler_GeomOptim}
\end{figure}

After the numerical design of the Si-slot waveguide in the linear and SA regime, we move on to the design of the coupler. We assume the two waveguides are at a sub-$\mu$m distance, measured by the gap ($g$) between their inner Si-ridge walls, Fig.~\ref{fig:Coupler_GeomOptim}(a), which leads to weak coupling for the tightly confining slot waveguides. Graphene sheets are omitted in these simulations as their effect is primarily absorptive and we are interested in the ideal, \textit{synchronized} lossless coupler. We extract the coupling length from the difference in phase constant of the two $x$-polarized (quasi-TE) supermodes of the coupler, the symmetric and anti-symmetric one, using a FEM-based mode solver. Figures~\ref{fig:Coupler_GeomOptim}(b) and (c) present the geometric and frequency dispersion of the coupling length, respectively; the latter is the primary parameter affecting the bandwidth of the device, which will be quantified in Section~\ref{sec:3:4:Bandwidth}. We also quantify the effect of nm-sized geometric deviations in the critical parameters of the coupler: the Si-ridge width ($w$), the air-slot size ($s$), and the gap between the two waveguides ($g$); the air-slot offset has a larger effect on the coupling length and it would be the critical feature in a fabricated device.

\subsection{Performance in CW Regime} \label{sec:3:3:CW}

To assess the nonreciprocal device performance, we start from the CW regime, where a harmonic signal at $\lambda_0=1550$~nm excites one of the coupler ports at various input powers. In terms of the coupled-equation system integrated to extract the output transmission, we use the simplified system of \eqref{eq:CMT_simple} where SA is the only nonlinear mechanism; the Kerr effect, in conjunction to SA, will be addressed in Section~\ref{sec:Kerr_CW_Results}. The attenuation coefficient for the graphene-loaded Si-slot waveguide has been numerically calculated in the saturation curve of Fig.~\ref{fig:SingleWG_LossSatComp} as a function of the CW input power; in a simpler case, one could use the $1/(1+\rho)$ phenomenological model applied directly to the attenuation coefficient, taking only the pair of $\alpha_0$ and $P_\mathrm{sat}$ values from the numerical solution, $\alpha=\alpha_0/(1+P_\mathrm{in}/P_\mathrm{sat})$. As explained in Section~\ref{sec:Coupled_NLSE}, this coupled-equation approach is valid under two justified approximations: (i) Graphene conductivity negligibly affects the phase constants of the waveguide modes, and thus their spatial profile, owing to the fact that $\mathrm{Im}\{\sigma^{(1)}\}$ is practically zero. (ii) We use single-polarization waveguides that form a coupler whose isolated modes have negligibly small spatial overlap, translating in very weak coupling. Consequently, all \textit{cross-nonlinear} parameters are very close to zero and can be safely excluded from the coupled system; the two equations correspond to the isolated graphene-loaded and unloaded waveguides, which are weakly coupled through $\kappa=\pi/2L_c$. 

\begin{figure}[] 
    \centering
    \includegraphics[]{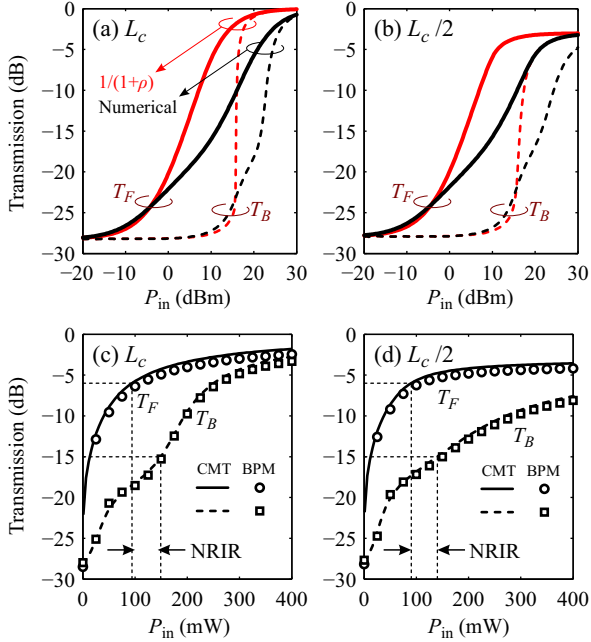}
    \caption{Forward (solid) and backward (dashed) transmission as a function of CW power. Panels (a) and (c) are for coupler length equal to $L_c$, while (b) and (d) are for $L_c/2$, with $L_c=600~\mu$m. Panels (a)-(b) compare the transmission curves for the phenomenological $1/(1+\rho)$ trend for the losses against the numerically calculated curve of Fig.~\ref{fig:SingleWG_LossSatComp}. Panels (c)-(d) compare the latter against NL-BPM simulation (markers).}
    \label{Fig:ResultsCW_SA}
\end{figure}

In order to attain a reasonably wide NRIR with realistic device footprint, we have numerically identified that a good choice is an asymmetric loading consisting of an uncoupled bilayer graphene ribbon, with low-power losses $\alpha_0=0.54$~dB/$\mu$m, and a coupling length of $L_c=600$~$\mu$m, realised by a gap $g=880$~nm between the two Si-slot waveguides. This corresponds to a normalized $\alpha_0/\kappa\approx48$ indicating that the device is far above the EP, owing to the large asymmetry in losses. We numerically integrate the coupled-equation system and extract the results for the CW case, presented in Fig.~\ref{Fig:ResultsCW_SA} as the forward and backward transmission against the input power, with black solid ($T_{F}$) and dashed ($T_{B}$) curves, respectively. In panels (a) and (b), the device length is equal to $L_c$ and $L_c/2$, respectively, which was found to exhibit approximately the same NRIR for the performance metrics selected, $T_{F}\ge-6$~dB and $T_{B}\le-15$~dB, corresponding to moderate forward insertion losses and adequate backward isolation, respectively. For these specifications, the nonreciprocal window spans from 100~mW to 160~mW, i.e., $\mathrm{NRIR}\approx2$~dB. With the saturation power of 0.22~mW, the normalized input powers that delimit the NRIR are approximately $[430,700]P_\mathrm{sat}$ i.e., far above the SA threshold. This can be explained by the relatively slow decrease of the numerically calculated SA curve, Fig.~\ref{fig:SingleWG_LossSatComp}, where an order of magnitude decrease in $\alpha$ happens 20~dB above $P_\mathrm{sat}$. In Fig.~\ref{Fig:ResultsCW_SA}, we also show the transmission curves when using the $1/(1+\rho)$ phenomenological model for the losses, with red curves, clearly leading to more optimistic device performance, namely 4~dB larger NRIR and 10~dB lower power thresholds. This result is also in line with the corresponding saturation curve in Fig.~\ref{fig:SingleWG_LossSatComp}, which decreases more rapidly towards zero than the numerically calculated one. Another remark that can be extracted is that the upper power limit of the NRIR is very sharp for the $1/(1+\rho)$ model, indicating that the transition from the isolation (nonreciprocal) to the breakdown (quasi-reciprocal) regime is abrupt. Finally, we note that due to the nonlinear nature of the device an optimal length for the device can potentially be found between $L_c$ and $L_c/2$, for given $T_{F,B}$ and NRIR limits.

The coupled-equation results in the CW regime were corroborated by nonlinear full-vector 3D beam propagation method (BPM) simulations. The BPM is a spectral paraxial method using an implicit stepping algorithm to propagate a vector excitation from an input cross-section ($xy$-plane) along the optical axis ($z$), until its output cross-section, from where the transmission at each port can be extracted for an integrated device such as the coupler. The propagation is done assuming a fixed reference index for the envelope phase, typically corresponding to the effective index of the propagation medium. BPM is valid under the slowly-varying envelope approximation justified for $z$-invariant reflectionless structures, or when the variations along the $z$-direction are ``slow'' inside each step. Our BPM was implemented with higher order triangular finite elements in the cross-section \cite{Selleri2001} and an iterative wide-angle (multi-step) correction in conjunction with the Crank-Nicolson scheme in the propagation direction \cite{Tsilipakos2011}. The difference in the phase constant (real part of effective index) of the two isolated waveguide modes is very small due to $\mathrm{Im}\{\sigma^{(1)}\}\approx0$, so the BPM applicability is ensured despite the high index-contrast waveguides used. The material nonlinearity, i.e., the E-field-dependent index or conductivity perturbation, is locally applied before each step of the BPM algorithm. Iterative stabilization is performed in each step (usually 2-3 iterations are enough) to account for the nonlinear perturbations. The nonlinear effect considered in this work is the saturation of the graphene surface conductivity, \eqref{eq:sigma1SA}, but other effects can also be incorporated, such as self-acting third-order effects from complex tensorial $\chi^{(3)}$ and $\sigma^{(3)}$, or perturbations from coupled systems (e.g., thermal effects, optically generated carrier diffusion/drift in silicon or graphene, electro-optic effects, multi-channel effects etc.). The NL-BPM results are depicted with markers in Fig.~\ref{Fig:ResultsCW_SA}(c)-(d), and are very close to the coupled-equation system solution (curves), validating its use. In our BPM simulations of the structure in Fig.~\ref{fig:Schem1}, the cross-section $xy$-plane was finely meshed resulting in approximately $10^5$ degrees of freedom and the $z$-propagation step-size was in the order of $\lambda_0$.

\subsection{Bandwidth Estimation} \label{sec:3:4:Bandwidth}

\begin{figure}[b] 
    \centering
    \includegraphics[]{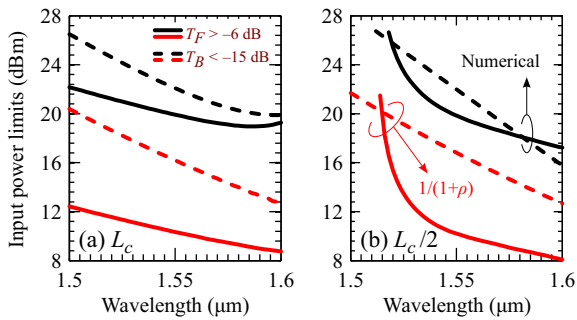}
    \caption{Forward (solid) and backward (dashed) input power limits for nonreciprocal CW operation vs. wavelength, accounting for coupling length dispersion. Panels (a) and (b) are for coupler length equal to $L_c$ and $L_c/2$, respectively, with fixed $L_c=600~\mu$m as calculated at 1550~nm. The NRIR is delimited between the solid and dashed lines of same color. Black and red curves correspond to the numerically calculated and the phenomenological model for the loss-saturation, respectively, Fig.~\ref{fig:SingleWG_LossSatComp}.}
    \label{Fig:BandwidthCW}
\end{figure}

In order to demonstrate the broadband nature of this device, we evaluate the NRIR across a 100~nm spectral window. Due to the broadband SA of graphene, the negligible imaginary part in its conductivity, and the symmetry of the Si-slot waveguides in the coupler, the main parameter defining the device bandwidth in the CW regime is the coupling length dispersion, Fig.~\ref{fig:Coupler_GeomOptim}(c). We numerically extract the threshold input powers for the previously used performance metrics, namely, $T_F \ge -6$~dB and $T_B \le -15$~dB, that delimit the NRIR. We analyzed both the full-length and half-length coupler, i.e., assuming device length equal to $L_c$ and $L_c/2$, respectively, with the corresponding results presented in the two panels in Fig.~\ref{Fig:BandwidthCW}. We also evaluated the NRIR dispersion both for the numerically extracted loss-saturation curve and the phenomenological curve $1/(1+\rho)$ that uses the low-power losses and the numerically extracted saturation power, Fig.~\ref{fig:SingleWG_LossSatComp}.

For the physically modeled full-length device [black curves in Fig.~\ref{Fig:BandwidthCW}(a)] we observe that the tolerable NRIR~$\approx2$~dB calculated for the central 1550~nm wavelength approximately covers a 70~nm band, and moreover improves to over 5~dB at lower wavelengths. This increase is due to the longer coupling length (smaller coupling coefficient) at lower wavelengths, which pushes the backward power threshold (``cross-saturation'' from the lossless waveguide excitation) higher than the forward threshold (``self-saturation'' from the SA waveguide excitation). The conclusion drawn here is that the bandwidth, like the NRIR, non-trivially depends on the saturation-curve of the waveguide, and an optimal component length can typically be found between $L_c/2$ and $L_c$, for the prescribed metrics.

For the half-length device, black curves in Fig.~\ref{Fig:BandwidthCW}(b), we find a narrower bandwidth as the NRIR closes entirely at $\pm30$~nm around the central wavelength. Note that when the forward power threshold (solid lines) curve crosses the backward threshold (dashed lines), we have the onset of an inversion of the directionality of the nonreciprocity; extracting the opposite metrics from the transmission curves (high $T_B$ and low $T_F$) can potentially unveil an \textit{opposite polarity} regime for the same isolator device. Finally, we observe once more the overly optimistic performance predicted by the phenomenological trend (red curves in Fig.~\ref{Fig:BandwidthCW}), leading to wider NRIR, lower power thresholds, and broader bandwidth.

\section{Further Considerations} \label{sec:4:Further}
\subsection{Performance in Pulsed Regime} \label{sec:Pulsed_Results}

The device performance can also be assessed in the pulsed regime, taking into account the frequency dispersion in the system, Eq.~\eqref{eq:NLSE_Dispersion}. In this work, the SA is assumed broadband and instantaneous, the Kerr effect is neglected (more details in Section~\ref{sec:Kerr_CW_Results}), and $v_g^{(1)} \approx v_g^{(2)}$ owing to $\mathrm{Im}\{\sigma^{(1)}\}=0$; so, we use only the single-waveguide dispersion parameters $\beta_{2,3}$ (GVD and TOD) and the coupling length dispersion. For the former, numerical simulations accounting for both waveguide and material (silicon and oxide) dispersion at $\lambda_0=1550$~nm were used to extract $\beta_2=+6.7$~ps$^2$/m and $\beta_3=-0.015$~ps$^3$/m; these parameters vary negligibly in the 100~nm window around 1550~nm. The full coupling length dispersion was directly plugged into the equation system at the frequency domain, using the data from Fig.~\ref{fig:Coupler_GeomOptim}(c); the dominant dispersion term is approximately +48~$\mu$m/THz.

\begin{figure}[] 
    \centering
    \includegraphics[]{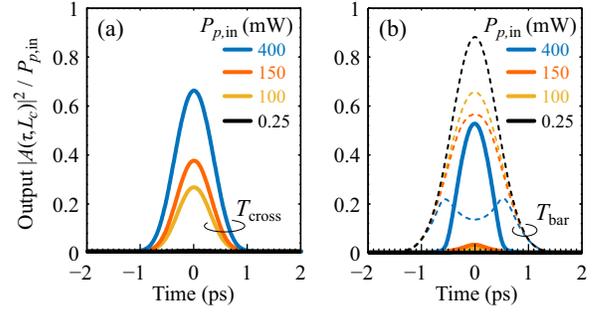}
     \caption{Normalized output pulses at the cross port of the coupler, for various peak-powers ($P_{p,\mathrm{in}}$), when exciting (a) the graphene-SA loaded waveguide in the forward direction, or (b) the lossless waveguide in the backward direction. The dashed curves in panel (b) correspond to the bar-port output.}
    \label{Fig:Pulsed}
\end{figure}

In the pulsed regime, the coupled NLSE system of \eqref{eq:NLSE_general} is integrated using the split-step Fourier method (SSFM), by driving a 1~ps FWHM pulse into the graphene-loaded or the unloaded waveguide port, at various peak powers. The normalized cross-transmitted pulses at the output of the $L_c$-long coupler are depicted in Fig.~\ref{Fig:Pulsed}, where we identify the trends predicted from the CW regime, without noticeable distortion. It is worth noting the twin-peak output pulse shape in the bar-ports when exciting the lossless waveguide, dotted curves in Fig.~\ref{Fig:Pulsed}(b), with peak powers above the NRIR: Only the central part of the pulse, that has sufficient power to saturate the graphene-loaded waveguide, is transmitted to the cross port. In this regime the device regresses to a quasi-reciprocal response, i.e., it has approximately the same cross-port transmission in both directions, e.g., the 400~mW curves in Fig.~\ref{Fig:Pulsed}. Finally, we estimate the onset of pulse distortion at 0.5~ps, primarily due to TOD (imparting an asymmetry in the temporal and spectral response) and secondarily due to GVD and/or coupling-length dispersion. This means that 1~ps pulses, requiring a bandwidth in the order of 10~nm, can be accommodated by the device whereas shorter pulses would require dispersion engineering.

\subsection{Third Order Effects in Graphene} \label{sec:Kerr_CW_Results}

Concerning the Kerr effect in integrated waveguides comprising graphene, various implementations, both theoretical and experimental, have revealed interesting phenomena, particularly in the perturbative regime, such as gate-tunable nonlinearity \cite{Alexander2018}. Kerr-induced nonreciprocity arises at more exaggerated power levels than SA (or it would necessitate higher graphene nonlinearity values), and usually relies on narrow-bandwidth (high-Q) resonators to further boost the nonlinear response \tdel{\nocite{Sounas2018_IEEE_AWPL}}\cite{DelBino2018}. In the $L_c$-long directional coupler, and in the absence of SA, the Kerr-induced nonreciprocity has an opposite isolation direction with respect to the SA-induced one: When exciting the graphene-loaded waveguide, the high Kerr effect (either focusing or defocusing) desynchronizes the coupler and inhibits coupling to the other waveguide, which leads to low cross-port transmission. When exciting the unloaded waveguide, coupling efficiency is not perturbed (coupler remains synchronized), so we have a high transmission; however, above a certain power threshold, phase modulation due to cross-coupling can desynchronize the coupler, leading, again, to low transmission and quasi-reciprocal response. In these cases, a geometric desynchronization of the waveguide coupler, e.g., different slot widths, can be used to tailor the response and reverse the polarity.

The third-order nonlinear effects were so far omitted to keep the focus on the SA phenomenon. Moreover, even though mathematically straightforward, incorporation of such effects in the coupled NLSE is physically not as simple, for a number of reasons: First and foremost is the free-carrier refraction, a non-perturbative process accompanying the photo-excited carrier induced SA, that has been shown to overshadow (perturbative) Kerr-type effects \cite{CastellLurbe2020,Vermeulen2018}. A second reason is the carrier-related nonlinearity coupled to the optical pulse propagation \cite{Mikhailov2019HEM}\tdel{\nocite{Soavi2019}}, whose implementation is complicated (both physically and computationally) but nonetheless important in the high-power regime. Thirdly, perturbative models typically predict that high values of $|\mathrm{Im}\{\sigma^{(3)}\}|$ (which contributes to the real part of $\gamma$) are attained for chemical potential close to the two-photon absorption resonance ($|\mu_c|\approx\hbar\omega/2$), with considerable dispersion, i.e., a tenfold (or more) decrease when frequency and/or $|\mu_c|$ are tuned away from that \cite{Cheng2015,Mikhailov2016Quantum}. Lastly, another point of caution is that solutions of perturbative models diverge at low-$|\mu_c|$ regions. Taking all these into account, and recalling that $|\mu_c|\rightarrow0$ is required for high-contrast (saturable) losses in graphene, reveals that including the Kerr effect in our proof-of-concept device should be done with caution and mainly in the direction of exploring other possible system dynamics. 

In this spirit, we explore Kerr-induced nonreciprocity in the nonlinear coupler, in the presence or absence of SA. To fully exploit the Kerr effect one should bias the graphene ribbon so that its chemical potential is above the half-photon energy, where graphene is almost transparent. Accurate expressions for interband and intraband linear monolayer conductivity at room temperature (quasi-equilibrium regime) reveal a local minimum of $\mathrm{Re}\{\sigma^{(1)}\}\approx\sigma_0/20$ at $|\mu_c|\approx0.55$~eV. Using this value for linear surface conductivity together with a defocusing value of {$\sigma^{(3)}=+j1.4\times10^{-21}$~S(m/V)$^2$} for the third-order nonlinear surface conductivity of a graphene monolayer \cite{Chatzidimitriou:20}, we can extract $\gamma=-44000$ and $+45$~m$^{-1}$W$^{-1}$ for the graphene-bilayer loaded and unloaded Si-slot waveguides, respectively, using the expressions in \cite{Chatzidimitriou:2015}; the nonlinear index of silicon is $n_2=2.5\times10^{-18}$~m$^2$/W. The rather high value for graphene $\gamma$ is due to the extremely high overlap of graphene with the $x$-polarized mode in the slot waveguide; it is worth pointing out that the maximization of $\gamma$ effectively coincides with the maximization of $a_0$, Fig.~\ref{fig:SingleWG_GeomOptim}, in the sense that they both depend on the graphene/E-field overlap (maximal light-matter interaction) in the waveguide cross-section. In this work, the real part of $\sigma^{(3)}$, related to perturbative SA or TPA (for negative or positive sign, respectively), is omitted as theoretical predictions show that it is generally lower and moreover exhibits a transition near half-photon energy \cite{Cheng2015,Mikhailov2016Quantum}. Moreover, we verified that Si-originating TPA and corresponding free-carrier absorption and refraction \cite{Pitilakis:2013} were negligible in the intensity ranges considered, due to the low E-field overlap with the Si ridges in the slot waveguide. 

\begin{figure}[] 
    \centering
    \includegraphics[]{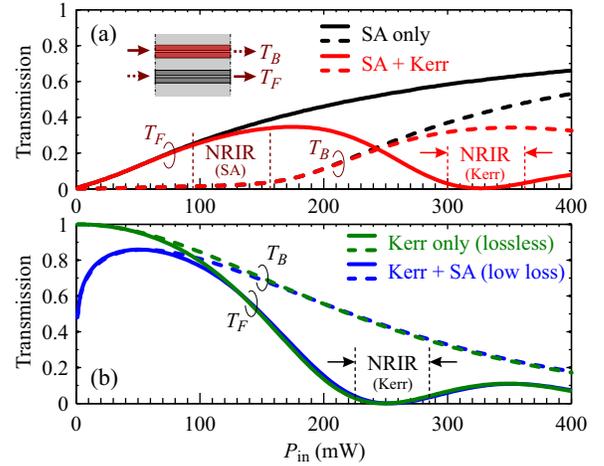}
    \caption{Forward and backward transmission as a function of input CW power. (a) SA only vs. SA+Kerr, where two nonreciprocal intensity ranges of opposite polarity open. (b) Kerr, with low saturable losses vs. hypothetical lossless case. The device length in all cases is $L_c=600~\mu$m.}
    \label{fig:KerrComparison_CW}
\end{figure}

Inserting the nonlinear parameters contributing to a power-dependent self-phase shift in each of the coupled equations [i.e., CW form of \eqref{eq:NLSE_general} with now asymmetric non-zero $\gamma^{(1,2)}$], we extract the forward and backward transmission curves against the input power, Fig.~\ref{fig:KerrComparison_CW}, for four scenarios:
(i) SA only with $\alpha_0=0.54$~dB/$\mu$m, (ii) Kerr and SA with $\alpha_0=0.54$~dB/$\mu$m, (iii) Kerr and SA now with low saturable losses, with $\alpha_0=0.027$~dB/$\mu$m, and (iv) a hypothetical lossless graphene configuration that only exhibits Kerr effect. 
In all scenarios the same SA curve \textit{shape} of Fig.~\ref{fig:SingleWG_LossSatComp} was assumed and $\gamma^{(1)}=-44000$ and $\gamma^{(2)}=+45$~m$^{-1}$W$^{-1}$, except in scenario (i) where $\gamma^{(1,2)}=+45$~m$^{-1}$W$^{-1}$. In Fig.~\ref{fig:KerrComparison_CW}(a), we observe that the combination of Kerr and SA opens two non-overlapping, equally-sized NRIR of opposite polarity, with the Kerr window appearing at three times higher power. When SA is diminished or switched off, Fig.~\ref{fig:KerrComparison_CW}(b), the NRIR opens slightly lower, and is well predicted by the nonlinear coupler theory \cite{Pitilakis:2013}, $P_\mathrm{in} \approx \pi\sqrt{3}/(\gamma^{(1)} L_c)$. Note that if non-saturable (background) graphene losses, e.g., due to intraband absorption, were considered in the Kerr-only scenario (iv), then the cross-transmission is considerably reduced in both directions; this is due to the very low $L_\mathrm{eff}\approx1/\alpha \ll L_c$. 

As a closing remark in this subsection, we note that the manifestation of third-order effects (such as self/cross phase/amplitude modulation, or four-wave mixing in general) is actually enabled by SA. The combination of SA and self-defocusing Kerr in graphene can give rise to interesting phenomena, such as soliton-like pulse compression in the normal-dispersion regime, $\beta_2>0$ \cite{Chatzidimitriou:20}. However, we stress that high-power illumination non-negligibly alters the 2D material and therefore its nonlinear parameters cannot be safely considered constant across so high power-contrast, especially in dynamic situations (fs-pulse regime), or when carrier thermodynamics are involved. \tdel{Nevertheless, we are aware that isolating one phenomenon from the others is also in a sense unphysical, as it negates the underlying interconnection between them. In this regard, an \textit{ab initio} approach to graphene nonlinearity, with self-consistent solution of all pertinent phenomena is sought, which is the subject of future work.}

\subsection{Future prospects} \label{sec:Future}

Two future steps are easily identified: Firstly, developing a theoretical model for this non-Hermitian system, which can be used for performance prediction rules, in-line with the present numerical study. Secondly, implementing a more elaborate physical description of graphene nonlinear response, incorporating photo-generated carrier effects, diffusion, and thermodynamic aspects; this will improve the accuracy of the predicted performance, possibly to more favorable metrics.

Further development on the present concept could also include re-engineering of the waveguides and coupler to optimize specific aspects of the nonreciprocal response: adjusted transmission limits ($T_{F,B}$), maximized NRIR and/or bandwidth, minimized footprint, dual/opposite isolation directions based on interplay between Kerr and SA, etc. \trev{All these amount to tailoring the saturation curve, i.e., how the \textit{material} properties are imprinted onto the \textit{waveguide mode} through light-matter interaction, acting on the imaginary part of the effective modal index across different power regimes. Apart from minor geometry tweaks, this tailoring can possibly be accomplished by introducing more elaborate features, e.g., multiple differently-biased graphene layers, longitudinally varying or patterned sheets, or bulk semiconductor/plasmonic materials}. More intricate pulsed-regime studies would include dispersion engineering for fs pulse-shaping, spectral broadening, or full-duplex operation. Finally, this component could be considered as part of more complex system such as a tunable cavity, ring laser, or a space-time modulated structure.

\section{Conclusions} \label{sec:5:conclusions}

We have proposed and numerically studied a proof-of-concept broadband nonreciprocal integrated device relying on a directional Si-photonic coupler asymmetrically loaded with graphene, a nonlinear 2D material exhibiting broadband SA even at low intensities. We adopted an instantaneous model for graphene SA to probe the limitations in the device and engineered the structure for sub-mW saturation power ($P_\mathrm{sat}$) in the graphene-loaded waveguide. We unveiled the non-Hermitian nature of the system and the underlying EP, and proposed a coupled-NLSE formulation for its analysis, using parameters rigorously extracted from a full-vector FEM-based mode solver; the validity of this formulation was checked against a nonlinear FEM-based full-vector BPM. Our results indicate a nonreciprocal window (NRIR) for 100~mW peak powers, with a high bandwidth of tens of nanometers, owing to the non-resonant nature of the structure. We also found that the NRIR (i) is much higher than the $P_\mathrm{sat}$ of the isolated waveguide, (ii) it non-trivially depends on the shape and steepness of the waveguide saturation curve, and (iii) cannot be safely estimated by phenomenological models for SA, such as $\alpha=\alpha_0/[1+(P_\mathrm{in}/P_\mathrm{sat})]$, which are over-optimistic. In conclusion, usable half-duplex isolator performance can be attained, provided that sufficient saturable losses and/or component length are available.


\bibliographystyle{IEEEtran}
\bibliography{IEEEabrv, JQE2021.bib}


\end{document}